\RequirePackage{fix-cm}
\documentclass[smallextended]{svjour3}       
\smartqed  
\usepackage{graphicx}
\usepackage{amsmath}
\usepackage{braket}
\usepackage{latexsym}
\usepackage{amssymb}
\usepackage{color}
\begin{document}

\title{Coherence and entanglement under three qubit cloning operations}


\author{Suchetana Goswami         \and
        Satyabrata Adhikari       \and
        A. S. Majumdar}


\institute{Suchetana Goswami \at
              S. N. Bose National Centre for Basic Sciences, Block JD, Sector-III, Salt Lake, Kolkata-700106, India \\
              \email{suchetana.goswami@gmail.com}           
           \and
           Satyabrata Adhikari \at
              Delhi Technological University, Shahbad Daulatpur, Main Bawana Road, Delhi-110042, India \\
              \email{tapisatya@gmail.com}
           \and
           A. S. Majumdar \at
              S. N. Bose National Centre for Basic Sciences, Block JD, Sector-III, Salt Lake, Kolkata-700106, India \\
              \email{archan@bose.res.in}
}

\date{Received: date / Accepted: date}

\maketitle

\begin{abstract}
Coherence and entanglement are the two most crucial resources for various quantum information processing tasks. Here, we study the interplay of coherence and entanglement under the action of different three qubit quantum cloning operations. Considering certain well-known quantum cloning machines (input state independent and dependent), we provide examples of coherent and incoherent operations performed by them. We show that both the output entanglement and coherence could vanish under incoherent cloning operations. Coherent cloning operations on the other hand, could be used to construct a universal and optimal coherence machine. It is also shown that under coherent cloning operations the output two qubit entanglement could be maximal even if the input coherence is negligible. Also it is possible to generate a fixed amount of entanglement independent of the nature of the input state. 

\end{abstract}

\section{Introduction}

Out of three non-local quantum correlations (such as enanglement~\cite{EPR_35}, steering~\cite{S_35_36}-\cite{JWD_07} and Bell-nonlocality~\cite{B_64}), entanglement is the most widely applied resource in the field of quantum information~\cite{NC_00}. Various manifestations of entanglement among discrete, continuous and hybrid physical variables have been studied in the context of applications in information theoretic protocols such as dense coding~\cite{dense}, teleportation~\cite{tele} and cryptography~\cite{crypto}. Investigations of the resource theory of entanglement have uncovered rich tenets~\cite{horod}, and some surprising features such as intra-particle entanglement~\cite{intra}. The connection of entanglement with other defining features of quantum theory such as the uncertainty principle has been rigorously examined~\cite{memory}.

Recently, quantum coherence~\cite{baumgratz} has come to be appreciated as one of the fundamental features of quantum theory. It has been  realized that coherence embodies basic quantumness responsible for superposition of quantum states, from which all quantum correlations arise in composite systems. As with entanglement, several measures have been suggested to quantify coherence~\cite{measures}. Interesting connections of coherence with thermodynamic properties of multipartite systems have been pointed out~\cite{thermo}. Efforts are on to develop resource theories of coherence enabling it to be used for detection of genuine non-classicality in physical states, and advantage in physical tasks over those performed using classical resources~\cite{resource}.

The relation of coherence with other resources in quantum theory forms an interesting arena of study. In a recent work, Streltsov {\it et al.}~\cite{streltsov} have provided an important insight into the linkage of coherence with entanglement. Based upon the observation that two-qubit incoherent operations can generate entanglement only when the input state is coherent, they have shown that the input state coherence provides an upper bound on the generated two-qubit entanglement. In another recent work, the complementarity of local coherence measures has been used to derive a nonlocal advantage of coherence in the form of enabling quantum steering~\cite{mondal}. In entanglement theory, it is known that the robustness (robustness of the state means here that the state does lose less quantum information in the quantum teleportation through noisy channels.) of GHZ and W states depends on the types of noisy channel \cite{JHJKY_08} while W state is more robust against qubit loss \cite{KCWHS_16}. In the resource theory of coherence, Y-Luo et. al. \cite{LLH} have shown that if one qubit is lost from GHZ state then the state will become incoherent but in case of W state, if one qubit is lost then the remaining two-qubit state remain coherent.  Moreover, they have defined  inequivalent classes of multipartite coherence states in the same spirit as in entanglement theory.
The connection between coherence and nonlocal resources such as entanglement is important to understand from both the perspective of quantum foundations and information theoretic applications, and thus deserves further study in various contexts.

In the present work we pose the question as to how the linkage between coherence and entanglement fares in the presence of additional parties or qubits. Specifically, we study the relationship between two-qubit entanglement and coherence under three-qubit operations. Quantum cloning provides a prototypical example of three-qubit operations, and here we employ coherent and incoherent cloning operations to investigate the connection between coherence and entanglement of the input and output states. For this purpose we consider different categories of cloning machines, such as the Wootters-Zurek~\cite{wootters} mechanism which acts as an incoherent operation, the Buzek-Hillary state independent cloning machine~\cite{buzek} which performs coherent operations. Also we consider phase covariant~\cite{BCDM_00} and state dependent universal~\cite{BDEFMS_98} cloning machines in order to undertake our study. Cloning could play an efficient role in resource replication, and in the present context we propose an optimal quantum coherence machine using our analysis.

The plan of the paper is as follows. In section $2$, we discuss three qubit incoherent operations which could lead to vanishing coherence and entanglement at the output. In section $3$, we study three qubit coherent operations and show how they could be used to construct a universal and optimal coherence machine that generates a fixed value of output coherence irrespective of the input state parameters. In section $4$, we investigate further the relation between coherence and entanglement in the context of coherent cloning operations. We find that maximally entangled two qubit output states could be generated even if the input state coherence is negligible.
Lastly in section $5$, we study the behaviour of coherence and entanglement of the output single party and two party states for state dependent cloning operations.  A summary of our main results is provided in section $6$.

\section{Entanglement and coherence in reduced two qubit system under incoherent quantum operations}

In this section we consider a three qubit incoherent quantum operation and investigate the coherence and entanglement generated in two qubit reduced state when third ancilla qubit is traced out. Coherence is an elementary property of quantum theory, which is basically a measure of quantumness  arising from the superposition principle of quantum mechanics. Also this is a basis dependent quantity and also it might exist in a single partite systems. Based on the superposition principle, an arbitrary state can be classified into two types: incoherent and coherent state. A state $\rho$ is said to be incoherent if it can be expressed in the form
\begin{eqnarray}
\rho=\sum_{i} \rho_{i}\ket{i}\bra{i} 
\label{incoherents}
\end{eqnarray}
where $\ket{i}$ represents a fixed reference basis of the state. Otherwise, it is said to be a coherent state. This definition holds not only for single qubit systems but also for higher dimensional quantum systems. Now, an incoherent quantum operation is defined as a completely positive trace preserving map which takes an incoherent state into another. There are different classes of incoherent operation as well~\cite{SAP_17}. Mathematically, an incoherent quantum operation $\Lambda$ can be written as
\begin{eqnarray}
\Lambda(\rho)=\sum_{l} K_{l}\rho K_{l}^{\dagger}
\label{incoherentqo}
\end{eqnarray}
where the operators $K_{l}$ are incoherent Kraus operators. There are different types of measures to quantify the amount of coherence in a given quantum state. In our present analysis, we will employ the $l_{1}$ norm measure~\cite{baumgratz} defined as
\begin{eqnarray}
C_{l_{1}}(\rho)=\sum_{i\neq j} |\rho_{ij}|
\label{l1norm}
\end{eqnarray}
On the other hand for the purpose of measuring entanglement, here we consider concurrence~\cite{CKW_00} of the quantum state (as it is sufficient for two qubit scenario) and for any two qubit state $\rho$ it is given by,
\begin{eqnarray}
C(\rho)=max\lbrace(\sqrt{\lambda_{1}}-\sqrt{\lambda_{2}}-\sqrt{\lambda_{3}}-\sqrt{\lambda_{4}}),0\rbrace
\label{conc2}
\end{eqnarray}
Where, $\lambda_{1}$, $\lambda_{2}$, $\lambda_{3}$, $\lambda_{4}$ are the eigenvalues of the matrix $\rho_{f}$, with $\rho_{f}=\rho . \tilde{\rho}$ ($\tilde{\rho}= (\sigma_{y}\otimes\sigma_{y}).\rho.(\sigma_{y}\otimes\sigma_{y})$).
In order to motivate our study, let us here briefly return to the case of two qubit incoherent operations discussed earlier by Streltsov {\it et al.}~\cite{streltsov}. Consider the tensor product of an input coherent state
\begin{eqnarray}
\ket{\psi}_{a}=c_{1}\ket{0}_{a}+c_{2}\ket{1}_{a},~~~~|c_{1}|^{2}+|c_{2}|^{2}=1
\label{input10}
\end{eqnarray}
with the ancilla state $\ket{0}_{b}$, and the state of the composite system is given by
\begin{eqnarray}
\ket{\Phi}_{ab}&=&\ket{\psi}_{a} \otimes \ket{0}_{b} \nonumber\\
&=&c_{1} \ket{00}_{ab}+c_{2} \ket{10}_{ab} 
\label{tp}
\end{eqnarray}
Now, a two qubit unitary CNOT operation which is given as,
\begin{eqnarray}
U_{CNOT}=\ket{00}\bra{00}+\ket{01}\bra{01}+\ket{10}\bra{11}+\ket{11}\bra{10}
\label{CNOT}
\end{eqnarray}
applied on $\ket{\Phi}_{ab}$, results in the two qubit state given by
\begin{eqnarray}
\ket{\phi}_{out}=c_{1}\ket{00}_{ab}+c_{2}\ket{11}_{ab}
\label{tp1}
\end{eqnarray}
One may now consider the following cases. If either $c_{1}=0$ or $c_{2}=0$, the input state (\ref{input10}) is incoherent and it remains an incoherent state even after the application of CNOT operation. Since the CNOT operation takes an incoherent state to another incoherent state and takes the set of incoherent basis $\lbrace \ket{00}, \ket{01}, \ket{10}, \ket{11} \rbrace$ into another $\lbrace \ket{00}, \ket{01}, \ket{11}, \ket{10} \rbrace$, it can be regarded as an incoherent operation. If both $c_{1}\neq0$ and $c_{2}\neq0$, the input state is a coherent state and the application of the CNOT operation on the tensor product (\ref{tp}) will generate an entangled state which basically reproduces the fact that to generate entanglement through an incoherent operation one has to start with a coherent state. In this case we find that the amount of entanglement generated is equal to the amount of coherence present in the input state. In general, it has been shown~\cite{streltsov} that the maximum entanglement generated by an incoherent operation is given by the amount of coherence present in the input qubit. In other words, a two-qubit incoherent operation generates entanglement, only if the input state has non-vanishing coherence. It is hence natural to ask the question if such a result can be extended to systems involving additional qubits. Note also, that the coherence of the two qubit input state $\ket{\Phi}_{ab}$ is equal to the coherence of the two qubit output state (\ref{tp1}) and it is given by $2|c_{1}||c_{2}|$. Thus, the coherence of the output state depends on the input state parameters. If, on the other hand, we trace out the second system, i.e., the mode $b$ from the two qubit system (\ref{tp1}),  the qubit in mode $a$ is left in an incoherent state. By generating entanglement through this incoherent operation one has to pay the price in terms of reducing the amount of coherence in the outputs $\rho_{A}$ and $\rho_{B}$ compared to that present in the input state \cite{streltsov}. In fact, in this case the single-qubit state at the output end is incoherent while we have started with a coherent state. In our subsequent analysis with three qubit operations, we investigate further this issue of the amount of coherence retained in the output states and its relation to the entanglement generated using three qubit operations.

Let us first consider the Wootters-Zurek cloning operation, which is a three qubit quantum operation expressed as \cite{wootters}
\begin{eqnarray}
\ket{0} \ket{0} \ket{0} \rightarrow \ket{0} \ket{0} \ket{0}
\label{WZ1}
\end{eqnarray}
\begin{eqnarray}
\ket{1} \ket{0} \ket{0} \rightarrow \ket{1} \ket{1} \ket{1}
\label{WZ2}
\end{eqnarray}
where the first ket vector represents the input state, the second ket vector represents the blank state in which the input state is to be copied and the third ket vector represents the machine state. It is clear from equations (\ref{WZ1}) and (\ref{WZ2}) that the cloning operation transforms incoherent input state into incoherent output state and also it takes the set of incoherent basis $\lbrace \ket{0}, \ket{1} \rbrace$ into another $\lbrace \ket{0}, \ket{1} \rbrace$, thus the above defined cloning operation is an example of a three qubit incoherent operation. Note that, in \cite{MY02} the WZ cloning machine has been studied for higher dimensional systems. From the transformation rule of this type of higher dimensional cloning machines, it is clear that it keeps an incoherent input state incoherent.

Now, if we take the input qubit to be coherent in nature, we see that this incoherent operation (\ref{WZ1}-\ref{WZ2}) does not generate entanglement between the input qubit and the blank qubit, when the ancillary machine state is traced out. Let us take the input qubit to be of the form
\begin{eqnarray}
\ket{\psi^{in}}=\alpha\ket{0}+\beta\ket{1},~~~~|\alpha|^{2}+|\beta|^{2}=1
\label{input1}
\end{eqnarray}
When the state (\ref{input1}) passes through the cloning transformation given in (\ref{WZ1}-\ref{WZ2}), the resulting two qubit state at the output end after tracing out the machine qubit is given by
\begin{eqnarray}
\rho^{out}_{12}=|\alpha|^{2}\ket{00}\bra{00}+|\beta|^{2}\ket{11}\bra{11}
\label{outputwz}
\end{eqnarray}
Also, the density operators of the copy qubits are given by
\begin{eqnarray}
\rho^{out}_{1}=\rho^{out}_{2}=|\alpha|^{2}\ket{0}\bra{0}+|\beta|^{2}\ket{1}\bra{1}
\label{copyoutputwz}
\end{eqnarray}
The following observations can be made from the equations (\ref{outputwz}-\ref{copyoutputwz}). It can easily be seen that the state described by the density operator $\rho^{out}_{12}$ is not entangled. Therefore, the transformation (\ref{WZ1}-\ref{WZ2}) is an example of a three qubit incoherent operation that does not generate entanglement between the input qubit and the blank qubit when machine qubit is traced out, even if we start with a coherent input state. We further find that the state described by the density operator $\rho^{out}_{12}$ is an incoherent two qubit state. The copy qubits generated at the output described by the density operators $\rho^{out}_{1}=\rho^{out}_{2}$ are incoherent states too. Now, the quality of copying of the cloning machine can be expressed in terms of the distance between the initial and the reduced copied state at the output end, measured by the Hilbert-Schmidt norm given as,
\begin{eqnarray}
D_a= Tr[(\rho^{in}-\rho^{out}_{1})^2]
\label{HS_norm}
\end{eqnarray}
In case of Wootters-Zurek cloning machine, one obtains the distance as,
\begin{eqnarray}
D_a= 2 |\alpha|^2 (1-|\alpha|^2)
\label{HS_WZ}
\end{eqnarray}
where, $D_a$ is known as the copy quality index. Averaging over all input states, one can obtain the copy quality as, 
\begin{eqnarray}
D_a=\frac{1}{3}
\label{HS_WZ1}
\end{eqnarray}

Note that, there is another cloning operation which does not generate entangelement at the output end, named phase covariant cloning \cite{BCDM_00} which can be regarded as an incoherent operation in single-qubit level (but generates coherence in two-qubit output) and it is given as,
\begin{eqnarray}
\ket{0}\ket{\Sigma}\ket{Q} &\rightarrow & [(\frac{1}{2}+\sqrt{\frac{1}{8}}) \ket{00}+ (\frac{1}{2}-\sqrt{\frac{1}{8}}) \ket{11}] \ket{\uparrow} \nonumber \\
&&+ \frac{1}{\sqrt{8}} (\ket{01}+\ket{10}) \ket{\downarrow} \\
\ket{1}\ket{\Sigma}\ket{Q} &\rightarrow & [(\frac{1}{2}+\sqrt{\frac{1}{8}}) \ket{11}+ (\frac{1}{2}-\sqrt{\frac{1}{8}}) \ket{00}] \ket{\downarrow} \nonumber \\
&&+ \frac{1}{\sqrt{8}} (\ket{01}+\ket{10}) \ket{\uparrow}
\label{phasecov}
\end{eqnarray}
Here, without any loss of generality we consider, $\ket{\Sigma}=\ket{0}$, $\ket{Q}=\ket{0}$, $\ket{\uparrow}=\ket{0}$ and $\ket{\downarrow}=\ket{1}$.
Likewise in this scenario, it is never possible to genrate any entanglement starting even from a coherent state, as one can check starting from a most general form of single qubit coherent state in computational basis, given in Eq. (\ref{input1}) with $\alpha \neq 0$ and $\beta \neq 0$. So this type of cloning machine is also not effective for the purpose of generating entanglement.
The above results motivate us to consider next three qubit quantum operations which may not be incoherent and can not only generate entanglement between the input qubit and the blank qubit when ancillary machine qubit is traced out, but also generate coherence in the copy qubits at the output.


\section{Optimal Universal Two-qubit Quantum Coherence Machine}

In this section we will consider the Buzek-Hillary (B-H) cloning operations~\cite{buzek} to see that there exist two classes of
three qubit coherent quantum operations that generate coherence in the reduced two qubit system. In the first class, the generated coherence depends on input state parameters, while in the second class, the coherence in reduced two qubit system does not depend on the input state parameters.

To begin with, let us consider the three qubit quantum operation given by~\cite{buzek}
\begin{eqnarray}
\ket{0}_{a} \ket{0}_{b} \ket{0}_{c} &\rightarrow & {\sqrt{\frac{2}{3}}} \ket{0}_{a} \ket{0}_{b} \ket{0}_{c} + \nonumber \\
&& {\sqrt{\frac{1}{6}}} (\ket{0}_{a} \ket{1}_{b}+\ket{1}_{a} \ket{0}_{b}) \ket{1}_{c}
\label{BH1}
\end{eqnarray}
\begin{eqnarray}
\ket{1}_{a} \ket{0}_{b} \ket{0}_{c} &\rightarrow & {\sqrt{\frac{2}{3}}} \ket{1}_{a} \ket{1}_{b} \ket{1}_{c} + \nonumber \\
&&{\sqrt{\frac{1}{6}}} (\ket{0}_{a} \ket{1}_{b}+\ket{1}_{a} \ket{0}_{b}) \ket{0}_{c}
\label{BH2}
\end{eqnarray}
The above transformation is a two-qubit coherent quantum operation as it takes an incoherent state to two-qubit coherent state. The above transformation is also known as the optimal state independent B-H cloning transformation in the $\{|0\rangle, |1\rangle\}$ basis. If we take the partial trace over the ancillary machine qubit $c$ at the output end of (\ref{BH1}) and (\ref{BH2}),  the corresponding reduced two qubit density operators are given by
\begin{eqnarray}
\rho_{ab}^{out1} & = & \frac{2}{3} \ket{00}\bra{00} + \frac{1}{6} (\ket{01}\bra{01}+\ket{10}\bra{01} \nonumber \\
&& +\ket{01}\bra{10}+\ket{10}\bra{10})
\label{outtwoBH1}
\end{eqnarray}
\begin{eqnarray}
\rho_{ab}^{out2} & = & \frac{2}{3} \ket{11}\bra{11} + \frac{1}{6} (\ket{01}\bra{01}+\ket{10}\bra{01} \nonumber \\
&& +\ket{01}\bra{10}+\ket{10}\bra{10})
\label{outtwoBH2}
\end{eqnarray}
The entanglement \cite{wootters1,connor} and coherence of the states $\rho_{ab}^{out1}$ and $\rho_{ab}^{out2}$ are equal and given by $\frac{1}{3}$. Thus, the B-H cloning machine generates a two qubit coherent state starting from an incoherent input qubit.

Let us next consider the input state $\ket{\psi^{in}}$ (\ref{input1}) with non-zero state parameters $\alpha$ and $\beta$. If we apply the optimal universal B-H cloning transformations given in (\ref{BH1}) and (\ref{BH2}) on $\ket{\psi^{in}}$,  the two qubit cloned state at the output end comes out to be of the form,
\begin{eqnarray}
\rho_{ab}^{out} & = & \frac{2}{3} |\alpha|^{2} \ket{00}\bra{00} + \frac{\sqrt{2}\alpha \beta^*}{3}\ket{00}\bra{+}+ \frac{\sqrt{2}\alpha^* \beta}{3}\ket{+}\bra{00}+\frac{1}{3}\ket{+}\bra{+} \nonumber \\
&& +\frac{\sqrt{2}\alpha \beta^*}{3}\ket{+}\bra{11}+ \frac{\sqrt{2}\alpha^* \beta}{3} \ket{11}\bra{+}+\frac{2}{3} |\beta|^{2} \ket{11}\bra{11}) 
\label{out1BH}
\end{eqnarray}
where $\ket{+}=\frac{1}{\sqrt{2}}(\ket{01}+\ket{10})$. The amount of coherence contained in the state described by the density operator $\rho_{ab}^{out}$ is given by $\frac{4(\alpha^* \beta+\alpha \beta^*)+1}{3}$. ($*$ denotes the corresponding complex conjugate) Thus, in case of the B-H quantum cloning machine, the generated cloned two qubit output is always coherent. It can be observed that the coherence of the state $\rho_{ab}^{out}$ depends on the state parameters  $\alpha$ and $\beta$ of the input. But, it should be noted that the concurrence of this two party output state turns out to be $\frac{1}{3}$, which is independent of the input state parameters. Also, it is quite interesting to note that this cloning machine generates a constant amount of entanglement starting from any single qubit input state. Hence it can be used as a source of constant entanglement.

We have seen that if we use this coherent quantum operation (\ref{BH1}-\ref{BH2}), the coherence of the reduced two qubit output state depends on the input state. It would be interesting to design a universal coherence transformation that transforms an arbitrary state $\ket{\Psi}_{ab}$ which may or may not be coherent, to a two qubit coherent state. We demand the transformation to be universal in the sense that the coherence of the two qubit output state should not be depending on the input state parameters. To construct such a coherence transformation, let us start with the most general form B-H quantum cloning transformation given by
\begin{eqnarray}
\ket{0}_{a} \ket{0}_{b} \ket{Q}_{c} &\rightarrow &
\ket{0}_{a} \ket{0}_{b} \ket{Q_{0}}_{c} + (\ket{0}_{a}
\ket{1}_{b} +\nonumber\\&&\ket{1}_{a} \ket{0}_{b})
\ket{Y_{0}}_{c} 
\label{CIBH1}
\end{eqnarray}
\begin{eqnarray}
\ket{1}_{a} \ket{0}_{b} \ket{Q}_{c} &\rightarrow &
\ket{1}_{a} \ket{1}_{b} \ket{Q_{1}}_{c} + (\ket{0}_{a}
\ket{1}_{b} +\nonumber\\&&\ket{1}_{a} \ket{0}_{b})
\ket{Y_{1}}_{c}
\label{CIBH2}
\end{eqnarray}
Unitarity of the transformation gives the relations
\begin{eqnarray}
_{c}\langle Q_{i}|Q_{i}\rangle_{c}+2_{c}\langle
Y_{i}|Y_{i}\rangle_{c} =1,~~~~i=0,1
\label{U1}
\end{eqnarray}
\begin{eqnarray}
_{c}\langle Y_{0}|Y_{1}\rangle_{c} =0,
\label{U2}
\end{eqnarray}
Let us further assume the following orthogonality relations between the machine state vectors:
\begin{eqnarray}
_{c}\langle Q_{i}|Y_{i}\rangle_{c}=0,~~~i=0,1
\label{O1}
\end{eqnarray}
\begin{eqnarray}
_{c}\langle Q_{0}|Q_{1}\rangle_{c}=0,
\label{O2}
\end{eqnarray}
First let us apply the cloning transformation given in (\ref{CIBH1}) and (\ref{CIBH2}) on an incoherent input state, say, $\ket{0} (\ket{1})$. At the output end, the coherence and concurrence of the final two party state (while the state of the ancillary system is traced out) turn out to be the same and it is given by, $2\mu$, where, $\mu$ is given by,
\begin{eqnarray}
_{c}\langle Y_{0}|Y_{0}\rangle_{c}=_{c}\langle
Y_{1}|Y_{1}\rangle_{c}=\mu
\label{not1}
\end{eqnarray}
Secondly, applying the cloning transformation (\ref{CIBH1}) and (\ref{CIBH2}) on $|\psi^{in}\rangle$ given by Eq.(\ref{input1}), and taking the partial trace over the ancillary machine qubit $c$, we obtain the cloned two qubit state described by the density operator
\begin{eqnarray}
\varrho_{ab}^{out}&=&|\alpha|^{2} (1-2\mu) |00\rangle\langle00| +
\alpha\beta^* \frac{\nu}{\sqrt{2}} |00\rangle\langle+| + \alpha^* \beta\frac{\nu}{\sqrt{2}} |+\rangle\langle00| \nonumber\\&&
+2\mu|+\rangle\langle+|+\alpha\beta^* \frac{\nu}{\sqrt{2}} |+\rangle\langle11| + \alpha^* \beta \frac{\nu}{\sqrt{2}} |11\rangle\langle+| \nonumber\\&&
+|\beta|^{2}(1-2\mu)|11\rangle\langle11|) 
\label{outgBH}
\end{eqnarray}
where $\mu$ is given as Eq.(\ref{not1}) and $\nu$ is given by
\begin{eqnarray}
&&_{c}\langle Y_{0}|Q_{1}\rangle_{c}= _{c}\langle
Q_{0}|Y_{1}\rangle_{c}=\frac{\nu}{2}
\label{not2}
\end{eqnarray}
Using the Schwarz inequality, the range of the parameters $\mu$, $\nu$ are given by
\begin{eqnarray}
0\leq\mu\leq\frac{1}{2}~~~\textrm{and}~~~
0\leq\nu\leq2\sqrt{\mu}\sqrt{1-2\mu}\leq\frac{1}{\sqrt{2}}
\label{si}
\end{eqnarray}
Now, the coherence of the state described by the density operator $\varrho_{ab}^{out}$ is given by
\begin{eqnarray}
C_{l_{1}}(\varrho_{ab}^{out})=2\mu+2 (\alpha^* \beta + \alpha \beta^*)\nu 
\label{coh1}
\end{eqnarray}
The quantity $C_{l_{1}}(\varrho_{ab}^{out})$ is input state independent if $\nu=0$. In this case Eq.(\ref{coh1}) reduces to $C_{l_{1}}(\varrho_{ab}^{out})=2\mu$.
The maximum value of $C_{l_{1}}(\varrho_{ab}^{out})$ can be obtained by putting $\mu=\frac{1}{2}$, which leads to
\begin{eqnarray}
C_{l_{1}}(\varrho_{ab}^{out})=1 
\label{coh1b}
\end{eqnarray}
For these particular values of $\mu$ and $\nu$, it can be seen that the concurrence of the two party state is  maximum. Also note that, the copy quality index in this case turns out to be $\frac{1}{18}$ which is much less than that of WZ cloning machine (hence better quality of cloning) and also independent of the input state parameter \cite{buzek}.\\
Eq. (\ref{coh1b}) is the evidence of the fact that the coherence present in the two qubit output state is optimal and independent of the input state parameters. Thus, we are successful in constructing a universal quantum coherence machine starting from the B-H quantum cloning machine. Particularly, the optimal universal quantum coherence transformation is given by
\begin{eqnarray}
|0\rangle_{a} |0\rangle_{b} |0\rangle_{c} &\rightarrow &
\sqrt{\frac{1}{2}}(|0\rangle_{a} |1\rangle_{b}+|1\rangle_{a}
|0\rangle_{b}) |0\rangle_{c} \label{CIT1}
\end{eqnarray}
\begin{eqnarray}
|1\rangle_{a} |0\rangle_{b} |0\rangle_{c} &\rightarrow &
\sqrt{\frac{1}{2}}(|0\rangle_{a} |1\rangle_{b}+|1\rangle_{a}
|0\rangle_{b}) |1\rangle_{c}\label{CIT2}
\end{eqnarray}
It is now clear that optimal universal quantum coherence transformation can be obtained from B-H quantum cloning transformation by choosing the machine vector in such a way that $\mu=\frac{1}{2}$ and $\nu=0$. Let us now ascertain how well the B-H copying machine with machine parameters $\mu=\frac{1}{2}$ and $\nu=0$ copies the input qubit described by the density operator
\begin{eqnarray}
\rho^{in}&=&\ket{\psi_{in}}\bra{\psi_{in}} \nonumber\\
&=&|\alpha|^{2}\ket{0}\bra{0}+\alpha \beta^* \ket{0}\bra{1}+ \alpha^* \beta \ket{1}\bra{0} 
+|\beta|^{2}\ket{1}\bra{1} 
\label{inputrho1}
\end{eqnarray}
Since the B-H quantum cloning machine is considered to be symmetric in nature, the copies of the input state at the output end of the copying machine are identical and are given by
\begin{eqnarray}
\rho_{a}^{out}=\rho_{b}^{out}&=&(|\alpha|^{2} (1-2 \mu)+2 \mu)|0\rangle\langle0|+\sqrt{2}\nu\alpha\beta^*|0\rangle\langle1|
\nonumber\\&&+\sqrt{2}\nu\alpha^*\beta|1\rangle\langle0| + (|\beta|^{2}(1-2\mu)+2 \mu)|1\rangle\langle1|
\label{outputrho1}
\end{eqnarray}
The distance between $\rho^{in}$ and $\rho_{a}^{out}/ \rho_{b}^{out}$ can be measured by the Hilbert Schmidt norm, given as
\begin{eqnarray}
D_{a}&=&Tr[(\rho^{in}-\rho_{a}^{out})^{2}]=Tr[(\rho^{in}-\rho_{b}^{out})^{2}]\nonumber\\&&
=2\mu^{2}(1-4|\alpha|^{2}|\beta|^{2})+2|\alpha|^{2}|\beta|^{2}(\nu-1)^{2}
\label{hs1}
\end{eqnarray}
Note that, for $\nu= 1-2 \mu$, the quality of copy for the B-H cloning machine becomes input state independent and particularly for $\mu=\frac{1}{2}$ and $\nu=0$, the distance $D_{a}$ reduces to
\begin{eqnarray}
D_{a}=\frac{1}{2}
\label{hs2}
\end{eqnarray}
Equation (\ref{hs2}) indicates that the quality of the copy also does not depend on the input state parameter. Therefore, for the cloning machine parameter $\mu=\frac{1}{2}$ and $\nu=0$, the B-H quantum cloning machine becomes an input state independent quantum cloning machine, but it should be noted that this cloning machine is not optimal in terms of quality of cloning.

\section{Generation of entanglement from coherent operations}

A quantum operation is said to be a coherent operation if it generates coherence even from an incoherent state. It has been already seen that under the application of any coherent operation it is possible to generate enatnglement starting even from an incoherent state. 
Unlike incoherent quantum operations (free operation in coherence resouce theory), we have seen that when a coherent operation (\ref{BH1}) acts on the tensor product of an incoherent input state and an incoherent ancilla  state, it generates entanglement in the two qubit reduced state when the ancilla state is traced out. We found that the amount of entanglement generated in the two qubit reduced state is equal to the amount of coherence in it.  This leads us to the following result.

\textbf{Result:} If we construct a coherent operation $\Lambda^{c}$ in such a way that it generates a two qubit mixed state of the form
\begin{eqnarray}
\rho_{AB}&=&
a|00\rangle\langle00|+b|01\rangle\langle01|+c|01\rangle\langle10|+c^{*}|10\rangle\langle01|\nonumber\\&&+d|10\rangle\langle10|+e|11\rangle\langle11|
\label{mixedstate}
\end{eqnarray}
when applied on the tensor product of an incoherent input state and an incoherent ancilla  state, the output entanglement and coherence are related by
\begin{eqnarray}
C(\rho_{AB})\leq C_{l_{1}}(\rho_{AB})
\label{result100}
\end{eqnarray}
where $C(\rho_{AB})$ is the concurrence of the output two qubit state and $C_{l_1}$ is the $l_1$ norm measure of coherence of the corresponding state.\\
\textbf{Proof:} It is known that the concurrence of the two qubit mixed entangled state (\ref{mixedstate}) is given by \cite{connor,bruss},
\begin{eqnarray}
C(\rho_{AB})= max \lbrace 0, 2(|c|-\sqrt{ae}) \rbrace \label{result102}
\end{eqnarray}
From (\ref{result102}), we find that
\begin{eqnarray}
C(\rho_{AB})\leq 2|c| \label{result103}
\end{eqnarray}
Again, the $l_1$ norm of coherence of the two qubit mixed state (\ref{mixedstate}) is given by
\begin{eqnarray}
C_{l_{1}}(\rho_{AB})= |c|+|c|= 2 |c|
\label{result104}
\end{eqnarray}
Using (\ref{result103}) and (\ref{result104}), we have
\begin{eqnarray}
C(\rho_{AB})\leq C_{l_{1}}(\rho_{AB})
\label{result105}
\end{eqnarray}
Hence proved. 

In this section, we consider a coherent operation in the form of B-H quantum cloning machine to study the entanglement structure of the two qubit output state. Depending on the (coherent/incoherent) nature of the input state, we analyze the entanglement structure of two qubit state at the output end of the cloning machine. First, let us consider the case when the input state to be cloned is an incoherent state which is either of the form $|0\rangle$ or
$|1\rangle$. When $|0\rangle$ goes through the cloning transformation given by Eqs.(\ref{CIBH1}) and (\ref{CIBH2}), the two qubit output state is given by
\begin{eqnarray}
\rho_{ab}^{out3} &=& (1-2\mu) |00\rangle\langle00| + 2\mu
|+\rangle\langle+|
\label{outtwoBH1gen}
\end{eqnarray}
It is clear that the concurrence of the two qubit state $\rho_{ab}^{out3}$ given by Eq.(\ref{outtwoBH1gen}) is non-zero and given by $2\mu$. A similar result can be obtained when the input state to be cloned is of the form $|1\rangle$. Therefore, the general B-H quantum cloning transformation generates an entangled two qubit cloned state when the input state is an incoherent. A maximally entangled state is generated when $\mu=\frac{1}{2}$. The structure of the cloning transformation that generates the maximally entangled state of two cloned copies out of the incoherent input  state is the same as the state independent quantum coherence transformation given by (\ref{CIT1}-\ref{CIT2}).

In the second scenario, let us consider that the input state to be cloned is a coherent state $|\psi_{in}\rangle$ given by (\ref{input1}). When the general B-H quantum cloning transformation is applied on $|\psi_{in}\rangle$ and tracing out the cloning machine state vector, the resulting two qubit state of two cloned copies is entangled. We study the entanglement of the output two qubit state for two cases.\\
\textbf{Case-I}: If we perform a state independent B-H quantum cloning transformation given by (\ref{CIBH1}) and  (\ref{CIBH2}) with $\nu=1-2\mu$, on any arbitrary coherent input state $|\psi_{in}\rangle$, the output state is given by
\begin{eqnarray}
\varrho_{ab}^{out}&=&|\alpha|^{2} (1-2\mu) |00\rangle\langle00| +
\alpha\beta^* \frac{1-2\mu}{\sqrt{2}} |00\rangle\langle+| + \alpha^*\beta\frac{1-2\mu}{\sqrt{2}} |+\rangle\langle00|\nonumber\\&&
+2\mu|+\rangle\langle+|+\alpha\beta^* \frac{1-2\mu}{\sqrt{2}}|+\rangle\langle11| + \alpha^* \beta\frac{1-2\mu}{\sqrt{2}} |11\rangle\langle+| \nonumber\\&&
+|\beta|^{2}(1-2\mu)|11\rangle\langle11|) \label{outgBH1}
\end{eqnarray}
We find that the generated two qubit cloned state is entangled and it is clearly evident from the plot given below.
\begin{figure}[!ht]
\resizebox{7cm}{4cm}{\includegraphics{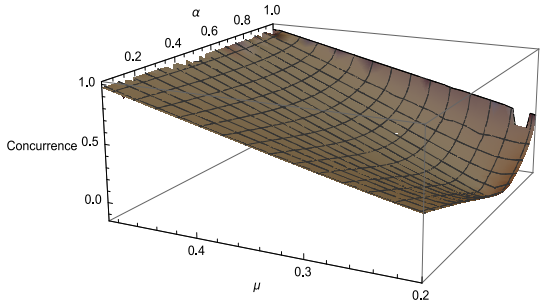}}
\caption{\footnotesize  Concurrence of the two qubit output state
is plotted against the machine parameter $\mu$ and the input state
parameter $\alpha$. } \label{conc_BH_1}
\end{figure}
From the plot, it can be seen that there exist state independent B-H quantum cloning transformations that cannot be used to generate two qubit entangled states. Additionally, one may note that the optimal state independent B-H quantum cloning machine can be used to generate a two qubit cloned state from a coherent input state. Also, one may observe that there exists a cloning transformation which generates maximum entanglement at the output end even when the coherence of the input is negligible.\\
\textbf{Case-II}: If we apply the optimal state independent quantum coherence transformation given by (\ref{CIT1}-\ref{CIT2}) on the coherent input state $|\psi_{in}\rangle$ (or may be on any incoherent input state, i.e. $|\psi_{in}\rangle$ either with $\beta=0$ or $\alpha=0$), then the two party output state is given by,
\begin{eqnarray}
\varrho_{ab}^{out}=\frac{1}{2}(|01\rangle\langle01|+|01\rangle\langle10|+|10\rangle\langle01|+|10\rangle\langle10|)
\label{optCOH_out}
\end{eqnarray}
The concurrence of this state is unity for any type of input state. So even for negligible amount of coherence it generates maximal entanglement.

\section{Study of coherence and entanglement in case of general state dependent cloning machine}

\begin{figure}[!ht]
\resizebox{7cm}{4cm}{\includegraphics{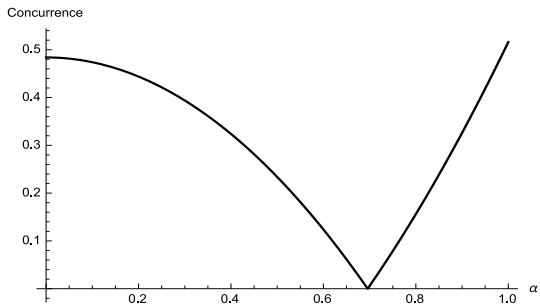}}
\caption{\footnotesize  Concurrence of the two qubit output state for general state dependent cloning machine
is plotted against the input stateparameter $\alpha$. } 
\label{statedep_conc}
\end{figure}

In this type of cloning \cite{BDEFMS_98}, the cloner operates a unitary operation on the composite Hilbert space of the three party state in the following way,
\begin{eqnarray}
\Ket{0} \Ket{0} \Ket{X} &\rightarrow & a \Ket{00} \Ket{A} + b_{1} \Ket{01} \Ket{B_{1}} + b_{2} \Ket{10} \Ket{B_{2}} + \nonumber \\
&& c \Ket{11} \Ket{C}
\label{statedep0}
\end{eqnarray}
\begin{eqnarray}
\Ket{1} \Ket{0} \Ket{X} &\rightarrow & \tilde{a} \Ket{11} \Ket{\tilde{A}} + \tilde{b_{1}} \Ket{01} \Ket{\tilde{B_{1}}} + \tilde{b_{2}} \Ket{10} \Ket{\tilde{B_{2}}} + \nonumber \\
&& \tilde{c} \Ket{00} \Ket{\tilde{C}}
\label{statedep1}
\end{eqnarray}
The above cloning operations are corresponding to the incoherent input states $\Ket{0}$ and $\Ket{1}$ respectively. Here the state $\Ket{X}$ represents the initial ancillary machine state and $\Ket{A}$, $\Ket{B_{1}}$, $\Ket{B_{2}}$, $\Ket{C}$, $\Ket{\tilde{A}}$, $\Ket{\tilde{B_{1}}}$, $\Ket{\tilde{B_{2}}}$, $\Ket{\tilde{C}}$ signify the ancillary machine state at the output end. As the operation of cloning is unitary, the coefficients in each case should satisfy the normalisation conditions,
\begin{eqnarray}
a^{2} + {b_{1}}^{2} + {b_{2}}^{2} + c^{2} = 1 \\
\tilde{a}^{2} + {\tilde{b_{1}}}^{2} + {\tilde{b_{2}}}^{2} + \tilde{c}^{2} = 1
\label{statedepnorm}
\end{eqnarray}
Here, we chose that $c=0$ and $\tilde{c}=0$ as the terms corresponding to these coefficients do not produce any productive output (neither it copies the state properly nor gives back the original state).\\
Now for our convinience let us choose (without any loss of generality), $\Ket{A}=\Ket{0}, \Ket{B_{1}}=\Ket{1}, \Ket{B_{2}}=\Ket{1}$ and hence $\Ket{\tilde{A}}=\Ket{1}, \Ket{\tilde{B_{1}}}=\Ket{0}, \Ket{\tilde{B_{2}}}=\Ket{0}$. Let us first start with an incoherent state $\ket{0}$ ($\ket{1}$), and under the transformation \ref{statedep0} (\ref{statedep1}) it can be observed that the single party cloned state at the output end is incoherent in nature. So, we can can call it as a single-qubit incoherent cloning operation. It is interesting to notice that under such type of cloning operation on these incoherent states, it is possible to generate an entangled state at the output end, whose concurrence is given by, $2 b_{1}$ (or, $2 \tilde{b_{1}}$) with $b_{1}=b_{2}$ (and, $\tilde{b_{1}}=\tilde{b_{2}}$)(assuming the cloning machine to be symmetric). Now let us consider the initial state to be coherent as given in, \ref{input1}. Now after passing through the cloning machine, the output three party state becomes, 
$\Ket{\psi_{out}} = \alpha [a \Ket{000} + b_{1} \Ket{011} + b_{2} \Ket{101}] + \beta [\tilde{a} \Ket{111} + \tilde{b_{1}} \Ket{100} + \tilde{b_{2}} \Ket{010}]$. Now to find the coherence of the copied state or the original state after the operation, one needs to trace out the two party state with respect to either the first party or with respect to the second party. For this cloning machine to be symmetric in nature, one should have the coherence of both the states to be equal. By assuming the symmetry of the operation one can get the coherence as $2(\tilde{a} b_{2}+ a \tilde{b_{2}}) \alpha\beta = 2(\tilde{a} b_{2}+ a \tilde{b_{2}}) \alpha \sqrt{1-{\alpha}^{2}}$. Now optimization with respect to $a, \tilde{a}, b_{1}, \tilde{b_{1}}$ gives that the maximum value of coherence of the final one party state for this type of cloning is $\sqrt{2}\alpha\beta$. The corresponding values of the parameters are $a=\frac{3}{4\sqrt{2}} = 0.695654$, $b_{1}= b_{2}= \frac{\sqrt{23}}{8}=0.507969
 $, $\tilde{a}= \frac{\sqrt{\frac{23}{2}}}{4}=0.718377$ and $\tilde{b_{1}}=\tilde{b_{2}}= \frac{3}{8}=0.491902$. One can easily see that compared to the state independent cloning the coherence of the final state is better in this case. Also, the entanglement also shows an improvement with respect to optimal BH cloning machine. The variation of concurrence with respect to the state parameter, for the optimized set of parameters of the machine is shown in \ref{statedep_conc} and it can be seen for most of the values of state parameter $\alpha$ the output two party state remains entangled. 

\section{Conclusions}

In this work we have considered three qubit cloning operations and studied
the two qubit output coherence and entanglement. The cloning operations have different copy quality indices and here we have considered cloning machines with different copying efficiency. In some cases it is independent of the input state parameters and for others, efficiency is dependent on the same. Recently, a bound has been obtained
on the two qubit entanglement in terms of the coherence of a single qubit input state
when an incoherent operation is performed on it~\cite{streltsov}. Our motivation for
the present study is to investigate further the connection between entanglement and coherence
in the context of cloning operations involving additional qubits. For this purpose
we have considered here two
types of well known cloning operations, {\it viz.}, the Wootters-Zurek copier~\cite{wootters},
and the Buzek-Hillery copier~\cite{buzek}. As we have discussed, our cloning operations
could be categorized into three qubit coherent and incoherent operations. We have
shown that the WZ cloning machine does not generate either coherence or entanglement
at the output. Cloning operations may be regarded as resource replicators in quantum information
processing. In the present work we next show that the BZ copier could act as a universal
coherence machine that generates a fixed amount of coherence in the two qubit output state
irrespective of the input state parameters.  Under the action of coherent cloning operations,
a relation
is obtained among the two qubit output coherence and entanglement. We have further shown that
under such operations, the
output entanglement could be maximal even if the input state coherence is negligible.

\vskip 0.3cm.

\end{document}